%% file: icassp21prasobh.tex
\documentclass{article}
\usepackage{spconf}
\usepackage{algorithm,graphicx}
\usepackage{algorithmic}
\usepackage{url}
\usepackage{cite}
\usepackage[usenames]{color}
\usepackage{amsfonts}
\usepackage{fancyhdr}
\usepackage{listings}%
\usepackage{arydshln}
\usepackage{amsmath,amssymb,amsthm}%
\usepackage{graphicx,psfrag,caption,subcaption}
\usepackage{listings}%
\usepackage{arydshln}
\usepackage{tikz}
\usetikzlibrary{shapes,arrows}

\usepackage{algorithmic}%
\input{macros}

\tikzstyle{block} = [draw, rectangle, 
    minimum height=3em, minimum width=4em]
 
\tikzstyle{sum} = [draw,  circle, node distance=1cm]
\tikzstyle{input} = [coordinate]
\tikzstyle{output} = [coordinate]
\tikzstyle{pinstyle} = [pin edge={to-,thin,black}]

\title{Millimeter Wave MIMO Channel Estimation with \\ 1-bit Spatial Sigma-delta Analog-to-Digital Converters}
\name{R.S. Prasobh Sankar and  Sundeep Prabhakar Chepuri 
}
\address{Indian Institute of Science, Bangalore, India\\
}

\begin{document}
\ninept
\maketitle

\begin{abstract}
This paper focuses on channel estimation for mmWave MIMO systems with 1-bit spatial sigma-delta analog-to-digital converters~(ADCs). The channel estimation performance with 1-bit spatial sigma-delta ADCs depends on the quantization noise modeling. Therefore, we present a new method for modeling the quantization noise by leveraging the deterministic input-output relation of the 1-bit spatial
sigma-delta ADC.   Using this new noise model,  we propose an algorithm for channel estimation for a  narrowband single-user mmWave line-of-sight MIMO system by determining the unknown angles and path attenuation that characterize the flat fading channel. Through simulations, we demonstrate that the performance of the developed method is comparable to the traditional analog systems and significantly better than the conventional 1-bit quantized systems.
\end{abstract}
\begin{keywords}
Channel estimation, direction estimation, mmWave MIMO, 1-bit quantization, spatial sigma-delta ADC.
\end{keywords}
\vspace*{-1mm}
\maketitle

\section{Introduction} \label{sec:intro} 

Nowadays, we are witnessing an unprecedented rise in the demand for higher data rates owing to the paradigm shift in our day-to-day life where we are more dependent on online resources than ever before. This motivates for faster deployment of mmWave MIMO systems, which promise data rates of the order of Gigabit per second~\cite{rappaport2013millimeter,ghosh2014millimeter}. In massive MIMO systems, the conventional way of using high-resolution quantizers per radio frequency (RF) chain can be expensive. To reduce the hardware costs, transceiver architectures with 1-bit ADCs~\cite{li2017channel} and hybrid beamformers~\cite{alkhateeb2014channel} are popularly suggested choices. 

In this paper, we focus on transceiver architectures with 1-bit ADCs for MIMO channel estimation. We particularly focus on a setup with 1-bit sigma-delta ($\Sigma\Delta$) modulator ADCs. Channel estimation with coarsely quantized data is challenging due to the non-linearity and noise introduced by the low-resolution ADCs. To obtain an equivalent linear model of the conventional 1-bit quantizer, usually the so-called {\it Bussgang decomposition}~\cite{bussgang1952crosscorrelation} is used~\cite{li2017channel,jacobsson2017throughput,demir2020bussgang}. In~\cite{li2017channel}, using the Bussgang decomposition, a linear minimum mean squared error (LMMSE) channel estimator was proposed for 1-bit massive MIMO systems, in which single antenna mobile users (MS) communicate with a multi-antenna base station (BS). Alternatively, instead of relying on the Bussgang decomposition, one can directly estimate the channel from the non-linear data model by solving a non-convex optimization problem. For a MIMO setup with multiple antennas at both the MS and BS, an optimization technique to alternately recover the amplitude from 1-bit measurements and estimate the channel was proposed in~\cite{qian2019amplitude}.

$\Sigma\Delta$ modulation is a well-known quantization technique that achieves higher effective resolution by oversampling and shaping the quantization noise to higher \textit{temporal} frequencies~\cite{aziz1996anoverview}.
Although $\Sigma\Delta$ ADCs are common for encoding temporal signals, they have been extended to encode spatial  signals as well~\cite{corey2016spatial,barac2016spatial}. Like the time domain, the shaping of the quantization noise to higher spatial frequencies is achieved via spatial oversampling and by integrating the quantization noise across the antennas~\cite{corey2016spatial,barac2016spatial}. MIMO transceiver architectures with low-resolution (including 1-bit) spatial $\Sigma\Delta$ ADCs have been used for interference cancellation~\cite{venkateswaran2011multichannel}, precoding~\cite{shao2019onebit}, and channel estimation~\cite{rao2019massive}. 

Recently, {\it unstructured} channel estimation based on a Bussgang-like decomposition in massive MIMO systems employing low resolution spatial $\Sigma\Delta$ ADCs has been considered in~\cite{pirzadeh2020spectral,rao2019massive,rao2020massive}. Assuming that the covariance structure of the channel between mutiple single antenna MSs and the BS equipped with multiple antennas is known, an LMMSE channel estimator was proposed~\cite{rao2019massive,rao2020massive}. Due to the underlying feedback structure in a $\Sigma\Delta$ ADC, computing the Bussgang decomposition in closed form is not possible. Hence, a recursive method~\cite{rao2019massive} to compute the Bussgang decomposition and an elementwise Bussgang decomposition per antenna~\cite{pirzadeh2020spectral}, both of which assume to know the covariance structure, have been proposed. In mmWave MIMO systems, (parametric) \emph{ angular channel} models are commonly used. For an angular channel model, which is parameterized by the angles of arrival (AoA) and angle of departure (AoD) of different paths and their gains,  it is not reasonable to assume that the channel covariance structure is known beforehand as knowing the channel covariance amounts to knowing the angles.  

 In \cite{gray1989quantization}, it is shown that the quantization noise in a \textit{temporal} 1-bit $\Sigma\Delta$ ADC is a deterministic function of the input and also provides closed-form expressions for the quantization noise. In this paper, we use this result to derive an analogous model describing the operation of a \textit{spatial} $\Sigma\Delta$ ADC, where we show that the quantization noise for antennas with large spatial indices can be considered to be uniformly distributed and uncorrelated to the corresponding input.
Next, using the developed noise model, we estimate the channel matrix in a mmWave MIMO system equipped with a 1-bit spatial $\Sigma\Delta$ precoder at the transmitter and a spatial 1-bit $\Sigma\Delta$ ADC at the receiver. The line-of-sight (LoS) MIMO channel is then estimated by finding the AoA and AoD using traditional MUSIC \cite{vantrees2002optimum} followed by estimating the complex path gain using least squares. From numerical simulations, we demonstrate that the developed method's performance for LoS MIMO channel estimation using 1-bit spatial $\Sigma\Delta$ ADCs is comparable to the traditional analog systems and significantly better than the 1-bit quantized systems. 

\vspace*{-2mm}
\section{System model} \label{sec:sysmodel}
In this paper, we consider a single-user uplink mmWave MIMO (SU-MIMO) system comprising of a BS and MS having uniform linear arrays (ULAs) with $N_{\rm{r}}$ and $N_{\rm{t}}$ antennas, respectively. Both the MS and BS are assumed to be equipped with first-order 1-bit spatial $\Sigma\Delta$ ADC having $N_{\rm t}$ and $N_{\rm r}$ channels, respectively.

\vspace*{-4mm}
\subsection{Spatial $\Sigma\Delta$ ADCs} \label{subsec:SpatialSD}

\begin{figure}
    \centering
    \includegraphics{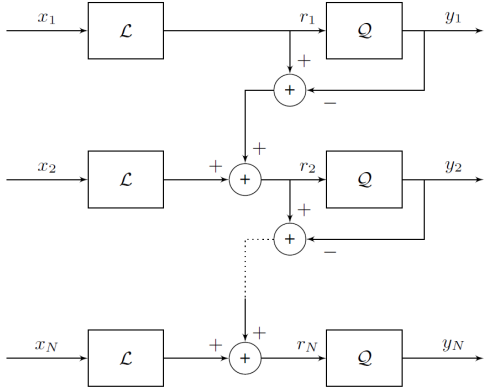}
   \caption{First-order, 1-bit spatial $\Sigma\Delta$ ADC. $\mathcal{Q}$ denotes 1-bit quantization and $\mathcal{L}$ denotes amplitude limitation.}
    \label{fig2}
    \vspace*{-5mm}
\end{figure}

Consider a first-order, 1-bit spatial $\Sigma\Delta$ ADC with $N$ channels ($N = N_{\rm{r}}$ at the BS and $N = N_{\rm{t}}$ at the MS) as illustrated in Fig.~\ref{fig2}. In a spatial $\Sigma\Delta$ ADC, the difference between the input and output of each quantizer (referred to as the quantization noise) is fed to the input of the quantizer in the next stage, forming an architecture identical to that of a temporal $\Sigma\Delta$ modulator, across space rather than time. Let us denote the input to the $\Sigma\Delta$ ADC at any given time instance as $\vx =  [x_{1},x_{2}, \ldots,  x_{N} ]\rT$. Then the output, $\mathbf{y} =  [ y_{1}, y_2, \ldots, y_{N}]\rT$,  of the $\Sigma\Delta$ ADC can be expressed as~\cite{rao2019massive, venkateswaran2011multichannel} 
\begin{equation} \label{basic_1}
    \mathbf{y} = \mathcal{Q}({\mathbf{U}\mathbf{x} - \mathbf{V}\mathbf{y}}),
\end{equation}
where $\mathbf{U}$ is the $N \times N$ lower triangular matrix with all ones in the lower triangular part 
and $\mathbf{V} = \mathbf{U} - \mathbf{I}_{N}$. Here, $\mathbf{I}_{N}$ denotes the $N \times N$ identity matrix, and $\mathcal{Q}(.)$ denotes a 1-bit quantizer having an output voltage level $b$, where the operation for any complex input $x$ is defined as  $\mathcal{Q}(x) = b \> {\rm sign}(\Re (x)) + j b \>  {\rm sign}(\Im (x))$.

We assume that the inputs are amplitude limited, i.e., $-b \leq \Re ({x_{i}}) , \Im ({ x_{i}}) \leq b, \> \forall \> i$. This assumption is commonly used with $\Sigma\Delta$ ADCs to prevent the quantization noise from growing unbounded because of the feedback~\cite{gray1989quantization, shao2019onebit}. This is ensured in practice using amplitude limiters $\mathcal{L}(.)$ as shown in Fig.~\ref{fig2}. Loss due to clipping operation can be minimized by selecting a sufficiently large $b$ so that we may write $\mathcal{L}(x_i) \approx x_i$ (see more details in Sec.~\ref{subsec:voltagelevel}).


\vspace*{-2mm}
\subsection{mmWave MIMO system model}

We consider a $N_{\rm{t}}$-channel $\Sigma\Delta$ ADC at the MS with an output level $b=1$. Let $\mathbf{T} \in \mathbb{C}^{N_{\rm{t}} \times M}$ be the input to the $\Sigma\Delta$ ADC at the MS, where each element is assumed to be amplitude limited (i.e.,
$-1 \leq \Re ({T_{i,j}}) , \Im ({ T_{i,j}}) \leq 1, \> \forall \> i,j$ ). The output of the $\Sigma\Delta$ ADC at the MS is denoted by $\mathbf{S} \in \{ \pm 1 \pm j \}^{N_{\rm{t}} \times M}$. With the total transmit power constrained to $P$, the symbols transmitted from the MS is given by
$\sqrt{\frac{P}{2N_{\rm{t}}}}\mathbf{S}$. Then the signal received at the BS, $\mathbf{X} \in \mathbb{C}^{N_{\rm{r}} \times M}$, can be written as
\begin{equation} \label{uq_received}
    \mathbf{X} = \sqrt{\frac{P}{2N_{\rm{t}}}}\mathbf{H}\mathbf{S} + \mathbf{W},
\end{equation}
where $\mathbf{H} \in \mathbb{C}^{N_{\rm{r}} \times N_{\rm{t}}}$ denotes the mmWave MIMO channel matrix, and $\mathbf{W} \in \mathbb{C}^{N_{\rm{r}} \times M}$ is the noise matrix. We assume that each entry of $\mW$ follows a complex Gaussian distribution with unit variance. 
Thus $P$ also denotes the uplink signal-to-noise ratio (SNR) of the SU-MIMO system.

This signal is then processed through the $N_{\rm{r}}$-channel $\Sigma\Delta$ ADC at the BS to obtain the received signal, which is denoted as
\begin{equation} \label{sd_eq2}
     \mathbf{Y} = \mathcal{Q}\left( \sqrt{\frac{P}{2N_{\rm{t}}}}\mathbf{U}\mathbf{H}\mathbf{S} + \mathbf{U}\mathbf{W} - \mathbf{V}\mathbf{Y}\right),
\end{equation}
where $\mathbf{Y} \in \mathbb{C}^{N_{\rm{r}} \times M}$. The columns of $\mathbf{Y}$ are vectors of the form in~\eqref{basic_1} correspond to different time instants.

The mmWave MIMO channels are often considered to be LoS or with a very few paths owing to the extensive pathloss exhibited at mmWave frequencies resulting in scenarios where the non-line-of-sight paths are too weak (when compared to the LoS path) for practical purposes  \cite{rappaport2013millimeter, ghosh2014millimeter}. In this work, we consider an LoS model for the MIMO channel matrix, which can be expressed as
\begin{equation} \label{ch_model1}
    \mathbf{H} = \alpha \mathbf{a}_{\rm{BS}}(\theta)\mathbf{a}_{\rm{MS}}\rH(\phi),
\end{equation}
where  $\alpha$, $\theta$, and $\phi$, denote the complex path gain, AoA at the BS and AoD at the MS, respectively. Here,  $\mathbf{a}_{\rm{MS}}(\phi)$ and $\mathbf{a}_{\rm{BS}}(\theta)$ denote the array steering vectors of ULAs at the MS and BS, respectively, and are given by
\begin{equation*}
     \mathbf{a}_{\rm{MS}}(\phi) =
    \begin{bmatrix}1, & e^{j(2\pi / \lambda) d \rm{sin}(\phi)}, & \cdots, & e^{j(N_{\rm{t}}-1)(2\pi / \lambda) d \rm{sin}(\phi) }  
    \end{bmatrix}\rT,
\end{equation*}
\begin{equation*}
     \mathbf{a}_{\rm{BS}}(\theta) =
    \begin{bmatrix}1, & e^{j(2\pi / \lambda) d \rm{sin}(\theta)}, & \cdots, & e^{j(N_{\rm{r}}-1)(2\pi / \lambda) d \rm{sin}(\theta) }  \end{bmatrix}\rT.
\end{equation*}
Here, $d$ is the inter-element sensor spacing in the ULAs at the MS and BS, and 
$\lambda$ is the signal wavelength. Without loss of generality, we assume that the first element is the phase reference of the array. In this paper, we consider the problem of pilot-based channel estimation, which involves estimation of the channel parameters $\theta$, $\phi$ and $\alpha$ from the output of the receiver  spatial $\Sigma\Delta$ ADC, $\mathbf{Y}$, given the pilots $\mT$ (input to the transmit spatial $\Sigma\Delta$ ADC). Before developing an estimator, we will model the quantization noise in a spatial $\Sigma\Delta$ ADC in the next section.

\vspace*{-2mm}
\section{Linearization and quantization noise} \label{sec:linquantnoise}

Typically, the non-linear quantization process is replaced with a linearized model, where the quantization operation is considered as the addition of a noise, which is assumed to be uniformly distributed and uncorrelated with the input of the quantizer. This assumption is usually satisfied only for certain scenarios, where an input has a sufficiently large dynamic range and the quantizer has a large number of levels. This means that the usual approach of  linearizing a quantizer will not provide much insights on the analysis of the quantization noise in 1-bit (spatial $\Sigma\Delta$) ADCs. In this section, we model the quantization noise in a first-order, 1-bit, spatial $\Sigma\Delta$ ADC. 

\subsection{Linearization of the 1-bit quantizer}

To begin with, let us re-visit the operation of the 1-bit spatial $\Sigma\Delta$ ADC with $N_{\rm r}$ antennas. For each antenna, we have
\begin{equation} \label{sdeqn1}
     y_{i} = \mathcal{Q}(r_{i}) = \mathcal{Q}\left( \sum_{l=1}^{i}x_{l} -\sum_{l=1}^{i-1} y_{l}\right), \> i=1,2,...,N_{\rm r}.
\end{equation}
We may linearize (\ref{sdeqn1}) and express the 1-bit quantization as the addition of an error term, $e_i$ , which we refer to as the quantization noise. That is, we can approximate $y_i$ as
\begin{equation} \label{sd_operation_scalar}
     y_{i} =  r_{i} + e_{i}, \> i=1,2,...,N_{\rm r},
\end{equation}
where we are not making any specific assumption on the distribution of $e_{i}$ and not assuming that $e_{i}$ is uncorrelated to $r_{i}$.
Collecting the outputs of different channels, we can approximate \eqref{basic_1} as
\begin{equation}  \label{sd_oper1}
    \mathbf{y} = \mathbf{r} + \mathbf{e},
\end{equation}
where $\mathbf{r} ={\mathbf{U}\mathbf{x} - \mathbf{V}\mathbf{y}} $. 
Then from\cite{gray1989quantization}, we can obtain a deterministic expression for the quantization noise, $e_{i}$, at each antenna in terms of the input as 
\begin{equation} \label{eq:q_noise}
   \Re( {e_{i}}) = b - (2b) \left<  \frac{(i-1)}{2} +\sum_{k=1}^{i} \frac{ \Re {( x_{k}})}{2b}  \right> \>,  \end{equation}
for $i = 1,2,\ldots,N_{\rm{r}},$ where $\left< . \right>$ denotes the fractional part function, which can be written in terms of the floor function as $\left< x \right> = x - {\rm floor}(x),\> \forall \> x \in \mathbb{R}$.
Substituting in (\ref{sd_oper1}), we get
\begin{equation} \label{q_noise2}
       \frac{1}{2b}\mathbf{U}\Re(\mathbf{y}) + \frac{1}{2}\mathbf{V}\boldsymbol{1} - \frac{1}{2}\boldsymbol{1}  = {\rm floor}\left(\frac{1}{2b}\mathbf{U}\Re(\mathbf{x}) + \frac{1}{2} \mathbf{V}\boldsymbol{1}\right),
\end{equation}
where $\boldsymbol{1}$ denotes the vector with all ones. Since the operation of the 1-bit quantizer for the real and imaginary parts of the input are identical, we also have
\begin{equation} \label{q_noise2i}
       \frac{1}{2b}\mathbf{U}\Im(\mathbf{y}) + \frac{1}{2}\mathbf{V}\boldsymbol{1} - \frac{1}{2}\boldsymbol{1}  = {\rm floor}\left(\frac{1}{2b}\mathbf{U}\Im(\mathbf{x}) + \frac{1}{2} \mathbf{V}\boldsymbol{1}\right).
\end{equation}


In essence, the equations (\ref{q_noise2}) and (\ref{q_noise2i}) provide a {\it deterministic} relation between the input and output of the 1-bit spatial $\Sigma\Delta$ ADC. 
\vspace*{-2mm}
\subsection{Linearization of the floor function}

Similar to the linearization of the 1-bit quantization in (\ref{sd_operation_scalar}), we can linearize the non-linear floor function and replace the floor function as the addition of an equivalent quantization noise, denoted by $\mathbf{q} \in~\mathbb{C}^{N_{\rm r}}$. Thus we have 
\begin{equation}
    {\rm floor}\left(\frac{1}{2b}\mathbf{U}\Re(\mathbf{x}) + \frac{1}{2} \mathbf{V}\boldsymbol{1}\right) = \frac{1}{2b}\mathbf{U}\Re(\mathbf{x}) + \frac{1}{2} \mathbf{V}\boldsymbol{1} + \Re (\mathbf{q}).
\end{equation}
Since $\mathbf{q}$ is a noise arising out of the floor operation, we know that the real (or imaginary) part of each entry of the vector $\mathbf{q}$ lies in the range~$[0,1)$. However, for antennas having larger indices (away from the phase reference), the dynamic range of  $[\frac{1}{2b}\mathbf{U}\Re(\mathbf{x}) + \frac{1}{2}\mathbf{V}\boldsymbol{1}]_{i}$ (or $[\frac{1}{2b}\mathbf{U}\Im(\mathbf{x}) + \frac{1}{2}\mathbf{V}\boldsymbol{1}]_{i}$) is large when compared to $[0,1)$. This means that for larger antenna indices, the operation of floor function is similar to that of a \emph{multi-level quantizer} with an input having a large dynamic range. So we may consider these noise terms $q_{i}$ to be uniformly distributed and uncorrelated with the corresponding inputs $x_{i}$ for antennas having larger indices. 

The uncorrelatedness of $q_{i}$ with $x_{i}$ for larger antenna indices is illustrated in Fig. \ref{Fig4} in Section 5.  However, we remark that this example does not prove that this assumption is always true irrespective of the voltage level $b$. As a simple counter example, if the voltage level $b$ is very large, the dynamic range of $[\frac{1}{2b}\mathbf{U}\Re(\mathbf{x}) + \frac{1}{2}\mathbf{V}\boldsymbol{1}]_{i}$ (or $[\frac{1}{2b}\mathbf{U}\Im(\mathbf{x}) + \frac{1}{2}\mathbf{V}\boldsymbol{1}]_{i}$)  can be small even for larger antenna indices. Nonetheless, with a careful choice of $b$ this assumption is usually satisfied. 

We can now re-arrange and combine (\ref{q_noise2}) and (\ref{q_noise2i}) to get
\begin{equation} \label{sd_final}
    \mathbf{y} = \mathbf{x} + (2b)\mathbf{U}^{-1}\Tilde{\mathbf{q}},
\end{equation}
where 
 $\tilde{\mathbf{q}} = (\Re(\mathbf{q}) + \frac{1}{2}\boldsymbol{1}) + j(\Im(\mathbf{q}) + \frac{1}{2}\boldsymbol{1}).$
Here, we may assume that $\Re(\Tilde{q}_{i}),\Im(\Tilde{q}_{i}) \sim {\rm uniform}(-0.5,0.5)$ for large values of $i$. 
Hence, from \eqref{sd_final}, we have a model describing the operation of the spatial $\Sigma\Delta$ ADC with a noise term, which is uncorrelated to the input $\mathbf{x}$ for antennas having larger indices.  In massive MIMO systems with large number of antennas, this assumption is reasonable for most of the antennas so that the term $(2b)\mathbf{U}^{-1}\Tilde{\mathbf{q}}$ can be viewed as a spatially high-pass filtered version of an i.i.d uniform random vector having a covariance matrix of $\frac{2b^2}{3}\mathbf{U}^{-1}\mathbf{U}^{-H}$. Thus shaping the noise to higher spatial frequencies.  In other words, we can use \eqref{sd_final} for channel or symbol estimation in MIMO systems with 1-bit $\Sigma\Delta$ ADCs.

\vspace*{-2mm}
\section{Channel estimation} \label{sec:chestm}
In this section, we present a channel estimation algorithm based on the traditional MUSIC \cite{vantrees2002optimum} after discussing the selection of the pilots.
\vspace*{-5mm}
\subsection{Pilot selection at the MS}

Recall that $\mathbf{S}$ is the output of the transmit $\Sigma\Delta$ ADC with $b=1$. Thus we have $\mathbf{S} \in \{ \pm 1 \pm j \}^{N_{\rm{t}} \times M}$. To simplify the channel estimation process, we desire $\mathbf{S}$ to be an orthogonal matrix with $\mathbf{S}\mathbf{S}\rH = \mathbf{S}\rH\mathbf{S} = 2N_{\rm{t}}\mathbf{I}_{N_{\rm{t}}}$. We also assume $N_{\rm t}$ to be a power of 2. We need to select the input to the transmit $\Sigma\Delta$ ADC, $\mathbf{T}$ so as to obtain the desired $\mathbf{S}$.
%
%

Let us express $\mathbf{S} = \mathbf{G} + j\mathbf{G}$ with $\mathbf{G}$ being the $N_{\rm{t}} \times N_{\rm{t}}$ Hadamard matrix.
Using equations (\ref{q_noise2}) and (\ref{q_noise2i}) that provide us a deterministic relationship between the input and output of a 1-bit $\Sigma\Delta$ ADC, and using the fact that $b=1$ for the transmit $\Sigma\Delta$ ADC, we can see that we need to select $\mathbf{T}$ such that 
\begin{eqnarray*}
     \frac{1}{2}\mathbf{U}{\mathbf{G}} + \frac{1}{2}\mathbf{V}\mathbf{1}\mathbf{1}\rT - \frac{1}{2}\mathbf{1}\mathbf{1}\rT  &= {\rm floor}(\frac{1}{2}\mathbf{U}\Re(\mathbf{T}) + \frac{1}{2} \mathbf{V}\mathbf{1}\mathbf{1}\rT);\\
     \frac{1}{2}\mathbf{U}{\mathbf{G}} + \frac{1}{2}\mathbf{V}\mathbf{1}\mathbf{1}\rT - \frac{1}{2}\mathbf{1}\mathbf{1}\rT  &= {\rm floor}(\frac{1}{2}\mathbf{U}\Im(\mathbf{T}) + \frac{1}{2} \mathbf{V}\mathbf{1}\mathbf{1}\rT).
\end{eqnarray*}
Clearly there are infinitely many choices of $\mathbf{T}$ that satisfy the above equations. By noting that ${\rm floor}(x) = x, \> \forall x \in \mathbb{Z}$, we see that one such choice would be to select $\mathbf{T}$ as
\begin{equation}
    \mathbf{T} = (\mathbf{G} - \mathbf{U}^{-1}\mathbf{1}\mathbf{1}\rT) + j(\mathbf{G} - \mathbf{U}^{-1}\mathbf{1}\mathbf{1}\rT),
\end{equation}
which results in the desired orthogonal matrix $\mS = \mG +j \mG$.

\subsection{Channel estimation at BS}
The channel estimation is performed in two stages, where we estimate the AoA and AoD in the first stage followed by the estimation of path gain in the second stage. Using the model that we have developed, recall that the output of the spatial $\Sigma\Delta$ ADC at the BS in \eqref{sd_eq2} can be written as 
\begin{equation} \label{eq_measurement}
    \mathbf{Y} = \sqrt{\frac{P}{2N_{\rm{t}}}}\mathbf{H}\mathbf{S} + \mathbf{N},
\end{equation}
where $\mathbf{N}$ is the sum of additive white Gaussian receiver noise $\mathbf{W}$ and the \textit{spatially high-pass filtered} quantization noise $2b\mathbf{U}^{-1}\Tilde{\mathbf{Q}}$ arising from the $\Sigma\Delta$ quantizer. Here, the columns of $\Tilde{\mathbf{Q}}$ contains vectors of the form $\Tilde{\mathbf{q}}$ in \eqref{sd_final}. The covariance matrix of the combined noise term (i.e., columns of $\mathbf{N}$) is given by 
\begin{equation}
    \mathbf{R}_{n} = \mathbf{I}_{N_{\rm{r}}} + \frac{2b^2}{3}\mathbf{U}^{-1}\mathbf{U}^{-H},
\end{equation}
where we assume that the receiver noise is independent from the quantization noise.
Since we know $\mR_n$, we first prewhiten \eqref{eq_measurement}. Then using the orthogonal matrix $\mathbf{S}$ (computed from $\mT$), we obtain a noisy version of the channel matrix
\begin{equation} \label{get_h1}
   \hat{\mH} = \mR^{-1/2}_n\mathbf{Y}\frac{1}{\sqrt{2PN_{t}}}\mathbf{S}\rH = \mR^{-1/2}_n\mathbf{H} + \hat{\mN},
\end{equation}
where $ \mR^{-1/2}_n$ is the pre-whitening matrix to whiten the noise $\mN$, and $\hat{\mN}= \mR^{-1/2}_n\mN\mS\rH/ \sqrt{2PN_t}$. 

\vspace*{-4mm}
\subsubsection{Estimation of the angles}
To obtain the AoA and AoD, we use the traditional MUSIC \cite{vantrees2002optimum}. Let us consider the singular value decomposition of the rank-1 channel $\hat{\mH} =  \mathbf{P}\mathbf{\Sigma}\mathbf{Q}\rH $. Since the noise term $\hat{\mN}$ is white and is uncorrelated with $\hat{\mH}$, we can compute the one-dimensional signal subspace due to the LoS path and noise subspaces as $\mathbf{P} = \begin{bmatrix} \mathbf{p}_{s} & \mathbf{P}_{n }\end{bmatrix}$ and $\mathbf{Q} = \begin{bmatrix} \mathbf{q}_{s} & \mathbf{Q}_{n }\end{bmatrix}$ such that 
 ${\mathcal{R}}(\mathbf{p}_{s}) = {\mathcal{R}}(\hat{\mH}) =  {\mathcal{R}}(\mathbf{R}^{-1/2}_n\mathbf{a}_{\rm{BS}}({\theta}))$  and ${\mathcal{R}}(\mathbf{q}_{s}) = {\mathcal{R}}(\hat{\mH}\rH)$ $\,\,=  {\mathcal{R}}(\mathbf{a}_{\rm{MS}}({\phi}))$, where $\mathcal{R}(.)$ denotes the range space.
 Here, $\mathbf{p}_{s} \in \mathbb{C}^{N_{\rm r} \times 1}$, $\mathbf{P}_{n} \in \mathbb{C}^{N_{\rm r} \times (N_{\rm r} - 1)}$, $\mathbf{q}_{s} \in \mathbb{C}^{N_{\rm t} \times 1}$, and $\mathbf{Q}_{n} \in \mathbb{C}^{N_{\rm t} \times (N_{\rm t} - 1)}$. 
 
 It is straightforward to observe that  $\mathbf{R}^{-1/2}_n\mathbf{a}_{BS}(\theta)$ and 
$\mathbf{a}_{MS}(\phi)$ are orthogonal to $\mathbf{P}_{n}$ and $\mathbf{Q}_{n}$, respectively. Hence we can estimate $\theta$ and $\phi$ from the location of the peaks of the MUSIC pseudo-spectra
\[
    \rho_{\rm BS}(\Tilde{\theta}) = \frac{1}{\|\mathbf{P}_{{n}}\rH \mathbf{R}^{-1/2}_n\mathbf{a}_{\rm{BS}}(\Tilde{\theta})\|^2_2} ; \,\,\,
    \>  \rho^{\rm MS}(\Tilde{\phi}) = \frac{1}{\| \mathbf{Q}_{{n}}\rH \mathbf{a}_{\rm{MS}}(\Tilde{\phi})\|^2_2},
\]
respectively.
\vspace*{-4mm}
\subsubsection{Path gain estimation} \label{subsec:pathgain}
Using the estimated AoA $\hat{\theta}$ and AoD $\hat{\phi}$, we can estimate the array manifolds $\mathbf{a}_{\rm{MS}}(\hat{{\phi}})$ and $\mathbf{a}_{\rm{BS}}(\hat{{\theta}})$.  From \eqref{ch_model1} and \eqref{get_h1}, we have
\begin{equation}
    \hat{\mH} = \alpha \mR^{-1/2}_n\va_{\rm{BS}}(\hat{\theta})\va_{\rm{MS}}\rH(\hat{{\phi}}) + \hat{\mN}.
\end{equation}
Then the least squares estimate of $\alpha$ is given by
\begin{equation}
    \hat{\alpha} = {\mathbf{d}\rH\hat{\mathbf{h}}}/{\| \mathbf{d}\|_2^{2}},
\end{equation}
where $\mathbf{d} = {\rm vec}(\mR^{-1/2}_n\va_{\rm{BS}}(\hat{\theta})\va_{\rm{MS}}\rH(\hat{{\phi}}))$ and $\hat{\mathbf{h}} = {\rm vec}(\hat{\mathbf{H}})$. Here, ${\rm vec}$ is the matrix vectorization operation. Hence, we can estimate the MIMO channel matrix as $\check{\mathbf{H}} = \hat{\alpha}\va_{\rm{BS}}(\hat{\theta})\va_{\rm{MS}}\rH(\hat{{\phi}})$.
\begin{figure}[!t]
    \centering
    \includegraphics[width = 0.75\columnwidth]{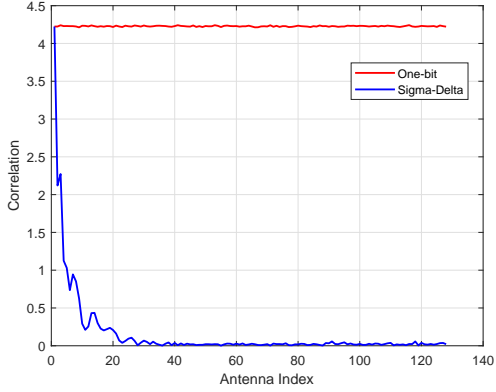}
    \caption{Correlation between the input and quantization noise.}
    \label{Fig4}
    \vspace*{-5mm}
\end{figure}

\vspace*{-2mm}
\subsection{Voltage level selection at the BS} \label{subsec:voltagelevel}
The choice of voltage levels is an important factor that determines the performance of MIMO systems employing $\Sigma\Delta$ ADCs. To predict the voltage levels, existing methods~\cite{pirzadeh2020spectral,rao2020massive} assume the knowledge of the channel covariance matrix, which is not realistic for angular channel models. 
A sub-optimal way to tackle this problem is to select the voltage level so that the effect due to clipping is minimized resulting in $\mathcal{L}(x_i) \approx x_i$. 

The input to $i$-th channel of the $\Sigma\Delta$ ADC, $x_{i}$, follows a unit variance complex Gaussian distribution centered around mean 
$\mu = \sqrt{\frac{P}{2N_{\rm t}}} \alpha e^{j 2\pi d (i-1)\sin(\theta)/\lambda}  \va_{MS}\rH(\phi){\vs}$, where it can be shown that $|\mu| \leq \sqrt{2PN_{\rm t}}$ for $|\alpha| = 1$. (Here, $\vs$ denotes a column of $\mathbf{S}$.) Since $\Re(x_j), \Im(x_j)$ each have variance $0.5$, a voltage level of $b = \sqrt{2PN_{\rm t}} + 3\sqrt{0.5}$  ensures that the clipping distortion is minimized for $\vert \mu \vert \leq \sqrt{2PN_{\rm t}}$. 

While this choice of $b$ works well for an LoS angular channel, the voltage level needs to be appropriately selected for other angular channel models with multipath components. We postponse this for future work.
\begin{figure}[!t]
    \centering
    \includegraphics[width=0.75\columnwidth]{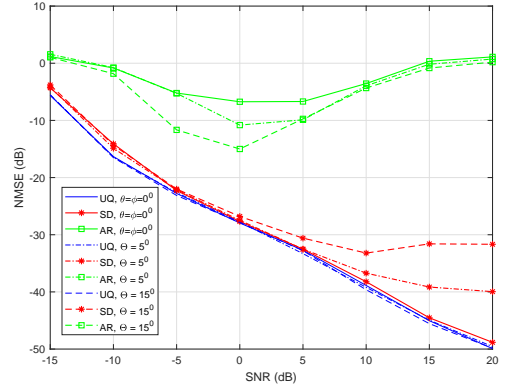}
    \caption{Channel estimation error.}
    \label{Fig_NMSE}
    \vspace*{-4mm}
\end{figure}

\vspace*{-3mm}
\section{Numerical simulations} \label{sec:numerical}

In this section, we compare the channel estimation performance of the proposed method \texttt{(SD)} with (a) an equivalent scheme applied on unquantized analog data \texttt{(UQ)}, which shall serve as the benchmark, and (b) \emph{amplitude retrieval} \texttt{(AR)} method~\cite{qian2019amplitude}. Since the methods in~\cite{rao2019massive,rao2020massive} assume the covariance structure at the input of the $\Sigma\Delta$ ADC, they are not suitable for angular channel models. We consider a setup with $N_{\rm{t}} = 8$, $N_{\rm{r}} = 128$, and a sensor spacing of $\lambda/8$.

The correlation between the input and quantization noise when $\theta = \phi = 0^{0}$ for an ${\rm SNR} = -5 {\rm dB}$ is shown in Fig. \ref{Fig4}, where we can see that the quantization noise for antennas away from the phase reference is uncorrelated to the inputs for a 1-bit spatial $\Sigma\Delta$ ADC, unlike its regular 1-bit counterpart as mentioned in Sec.~\ref{sec:linquantnoise}. 
 
The normalized mean squared error (NMSE), defined as $\mathbb{E}(\|\check{\mH} - \mH \|^2_F)/\mathbb{E}(\|\mH\|_{F}^2)$, of channel estimates under different scenarios are presented in Fig.~\ref{Fig_NMSE}. The results shown are computed by averaging over 200 independent Monte-Carlo realizations of the LoS MIMO channel where the AoA and AoD are selected at random from an angular sector having a width of $2\Theta$ centered around $0^\circ$, along with a randomly generated complex path gain with unit magnitude.  We can observe that the proposed method performs significantly better than the 1-bit MIMO channel estimation technique \texttt{AR} and is comparable to the unquantized case. However, due to the inevitable presence of the quantization noise at higher SNRs, the performance gap between \texttt{SD} and \texttt{UQ} increases, especially when the angular spread is large.

\vspace*{-2mm}
\section{Conclusions}
In this paper, we have proposed an angular channel estimation technique for mmWave massive SU-MIMO systems with 1-bit multichannel spatial $\Sigma\Delta$ ADCs at both the transmitter and receiver. The proposed algorithm does not require any prior knowledge of the channel correlation matrix. Through numerical simulations, we have demonstrated that our method performs significantly better as compared the conventional 1-bit MIMO channel estimation algorithm, reiterating the advantage of a $\Sigma\Delta$ ADC over its regular counterpart. 

\pagebreak

\bibliographystyle{IEEEtran}
\bibliography{IEEEabrv,references}

\end{document}

%% file: macros.tex
\def\cast{{
   \mathord{
      \hbox to 0em{
         \ooalign{
	   \smash{\hbox{$\ast$}}\crcr
	   \smash{\hskip-1pt\Large\hbox{$\circ$}} }
	 \hidewidth}
      \phantom{\bigcirc}
} }}

\def\bm#1{\mbox{\boldmath $#1$}}

\newcommand{\rH}{^{ \raisebox{1pt}{$\rm \scriptscriptstyle H$}}}
\newcommand{\rT}{^{ \raisebox{1.2pt}{$\rm \scriptstyle T$}}}

\newcommand{\bds}{\begin {itemize}}
\newcommand{\eds}{\end {itemize}}
\newcommand{\bdf}{\begin{definition}}
\newcommand{\blm}{\begin{lemma}}
\newcommand{\edf}{\end{definition}}
\newcommand{\elm}{\end{lemma}}
\newcommand{\bthm}{\begin{theorem}}
\newcommand{\ethm}{\end{theorem}}
\newcommand{\bprp}{\begin{prop}}
\newcommand{\eprp}{\end{prop}}
\newcommand{\bcl}{\begin{claim}}
\newcommand{\ecl}{\end{claim}}
\newcommand{\bcr}{\begin{coro}}
\newcommand{\ecr}{\end{coro}}
\newcommand{\bquest}{\begin{question}}
\newcommand{\equest}{\end{question}}

\newcommand{\larrow}{{\larrow}}

\newcommand{\argmin}{\ensuremath{\mathrm{arg}\min}}
\newcommand{\argmax}{\ensuremath{\mathrm{arg}\max}}

\newcommand{\va}{{\ensuremath{{\mathbf{a}}}}}

\newcommand{\vs}{{\ensuremath{{\mathbf{s}}}}}

\newcommand{\vx}{{\ensuremath{{\mathbf{x}}}}}

\newcommand{\mG}{{\ensuremath{\mathbf{G}}}}
\newcommand{\mH}{{\ensuremath{\mathbf{H}}}}

\newcommand{\mN}{{\ensuremath{\mathbf{N}}}}

\newcommand{\mR}{{\ensuremath{\mathbf{R}}}}

\newcommand{\mS}{{\ensuremath{\mathbf{S}}}}

\newcommand{\mT}{{\ensuremath{\mathbf{T}}}}

\newcommand{\mW}{{\ensuremath{\mathbf{W}}}}

\usepackage{latexsym}

\def\IC{\mathbb C}
\def\IN{\mathbb N}
\def\IZ{\mathbb Z}
\def\IR{\mathbb R}

\def\shat{^{\mathchoice{}{}%
 {\,\,\smash{\hbox{\lower4pt\hbox{$\widehat{\null}$}}}}%
 {\,\smash{\hbox{\lower3pt\hbox{$\hat{\null}$}}}}}}

\def\bSigma{{
      \ooalign{
      \smash{\hskip.4pt\raise.4pt\hbox{$\Sigma$}}\vphantom{}\crcr
      \smash{\hskip.7pt\raise.6pt\hbox{$\Sigma$}}\vphantom{}\crcr
      \smash{\hbox{$\Sigma$}}\vphantom{$\Sigma$}}
      \vphantom{\hbox{$\Sigma$}}
      }}
\def\bTheta{{
      \ooalign{
      \smash{\hskip.5pt\raise.5pt\hbox{$\Theta$}}\vphantom{}\crcr
      \smash{\hskip.0pt\raise.1pt\hbox{$\Theta$}}\vphantom{}\crcr
      \smash{\hbox{$\Theta$}}\vphantom{$\Theta$}}
      \vphantom{\hbox{$\Theta$}}
      }}
\def\bDelta{{
      \ooalign{
      \smash{\hskip.4pt\raise.4pt\hbox{$\Delta$}}\vphantom{}\crcr
      \smash{\hskip.7pt\raise.6pt\hbox{$\Delta$}}\vphantom{}\crcr
      \smash{\hbox{$\Delta$}}\vphantom{$\Delta$}}
      \vphantom{\hbox{$\Delta$}}
      }}
\def\bLambda{{
      \ooalign{
      \smash{\hskip.5pt\raise.5pt\hbox{$\Lambda$}}\vphantom{}\crcr
      \smash{\hskip.0pt\raise.1pt\hbox{$\Lambda$}}\vphantom{}\crcr
      \smash{\hbox{$\Lambda$}}\vphantom{$\Lambda$}}
      \vphantom{\hbox{$\Lambda$}}
      }}

\makeatletter

\def\bordermatrix#1{\begingroup \m@th
  \@tempdima 8.75\p@
  \setbox\z@\vbox{%
    \def\cr{\crcr\noalign{\kern2\p@\global\let\cr\endline}}%
    \ialign{$##$\hfil\kern2\p@\kern\@tempdima&\thinspace\hfil$##$\hfil
      &&\quad\hfil$##$\hfil\crcr
      \omit\strut\hfil\crcr\noalign{\kern-\baselineskip}%
      #1\crcr\omit\strut\cr}}%
  \setbox\tw@\vbox{\unvcopy\z@\global\setbox\@ne\lastbox}%
  \setbox\tw@\hbox{\unhbox\@ne\unskip\global\setbox\@ne\lastbox}%
  \setbox\tw@\hbox{$\kern\wd\@ne\kern-\@tempdima\left[\kern-\wd\@ne
    \global\setbox\@ne\vbox{\box\@ne\kern2\p@}%
    \vcenter{\kern-\ht\@ne\unvbox\z@\kern-\baselineskip}\,\right]$}%
  \null\;\vbox{\kern\ht\@ne\box\tw@}\endgroup}
\makeatother

\makeatletter
\def\argmin{\mathop{\operator@font arg\,min}}
\def\argmax{\mathop{\operator@font arg\,max}}
\makeatother

\def\bm#1{\mbox{\boldmath $#1$}}

\newcommand{\bbeta}{{\bm\beta}}

\newcommand{\bea}{\begin{array}}
\newcommand{\ena}{\end{array}}
\newcommand{\beq}{\begin{equation}}
\newcommand{\enq}{\end{equation}}

\newcommand{\beqa}{\begin{eqnarray}}
\newcommand{\enqa}{\end{eqnarray}}

\newcommand{\beqan}{\begin{eqnarray*}}
\newcommand{\enqan}{\end{eqnarray*}}

\newcommand{\AL}{\begin{enumerate}}
\newcommand{\ALE}{\end{enumerate}}

\def\addots{\mathinner{
    \mkern1mu\raise0pt\vbox{\kern7pt\hbox{.}}
    \mkern2mu\raise4pt\hbox{.}
    \mkern2mu\raise7pt\hbox{.}
    \mkern1mu}}

\def\sddots{\mathinner{
    \mkern.8mu\raise7pt\hbox{.}
    \mkern.8mu\raise4pt\hbox{.}
    \mkern.8mu\raise0pt\vbox{\kern7pt\hbox{.}}
    \mkern1mu}}

\def\saddots{\mathinner{
    \mkern.2mu\raise0pt\vbox{\kern7pt\hbox{.}}
    \mkern.2mu\raise4pt\hbox{.}
    \mkern.2mu\raise7pt\hbox{.}
    \mkern1mu}}

\def\sqplus{\mathbin{
	{\ooalign{\hfil\raise.3ex\hbox{\scriptsize
	+}\hfil\crcr\mathhexbox274\crcr\mathhexbox275}}
	}} 
\def\sqminus{\mathbin{
	{\ooalign{\hfil\raise.3ex\hbox{\scriptsize
	--}\hfil\crcr\mathhexbox274\crcr\mathhexbox275}}
	}}

\def\IC{{
   \mathord{
      \hbox to 0em{
	 \hskip-4pt
         \ooalign{
	   \smash{\hskip1.9pt\raise2.6pt\hbox{$\scriptscriptstyle |$}}\crcr
	   \smash{\hbox{\rm\sf C}} }
	 \hidewidth}
      \phantom{\hbox{\rm\sf C}}
} }}
\def\IN{
    {\ooalign{
   \smash{\hskip2.2pt\raise1.5pt\hbox{$\scriptscriptstyle |$}}\vphantom{}\crcr
   \hbox{\sf N}
	}}
	} 
\def\IZ{
    {\ooalign{
   \smash{\hskip1.9pt\raise0pt\hbox{$\sf Z$}}\vphantom{}\crcr
   \hbox{\sf Z}
	}}
	} 
\def\IR{
    {\ooalign{
   \smash{\hskip2.2pt\raise1.5pt\hbox{$\scriptscriptstyle |$}}\vphantom{}\crcr
   \smash{\hskip2.2pt\raise3.3pt\hbox{$\scriptscriptstyle |$}}\vphantom{}\crcr
   \hbox{\sf R}
	}}
	} 

\DeclareMathAlphabet{\mathcmb}{OT1}{cmr}{b}{n}

\def\bSigma{\ensuremath{\mathcmb{\Sigma}}}
\def\bLambda{\ensuremath{\mathcmb{\Lambda}}}

\def\bTheta{\ensuremath{\mathcmb{\Theta}}}

\newcommand{\SI}{\begin{indlist}}
\newcommand{\EI}{\end{indlist}}

\newcommand{\DL}{\begin{dashlist}}
\newcommand{\DLE}{\end{dashlist}}

\makeatletter
\def\setboxz@h{\setbox\z@\hbox}
\def\wdz@{\wd\z@}
\def\boxz@{\box\z@}
\def\underset#1#2{\binrel@{#2}%
  \binrel@@{\mathop{\kern\z@#2}\limits_{#1}}}
\def\binrel@#1{\begingroup
  \setboxz@h{\thinmuskip0mu
    \medmuskip\m@ne mu\thickmuskip\@ne mu
    \setbox\tw@\hbox{$#1\m@th$}\kern-\wd\tw@
    ${}#1{}\m@th$}%
  \edef\@tempa{\endgroup\let\noexpand\binrel@@
    \ifdim\wdz@<\z@ \mathbin
    \else\ifdim\wdz@>\z@ \mathrel
    \else \relax\fi\fi}%
  \@tempa
}
\let\binrel@@\relax%
\makeatother
